\documentclass[twocolumn]{aa}
\usepackage{graphicx}
\usepackage{txfonts}
\def\ltsim{\hbox{\raise 2pt \hbox {$<$} \kern-1.1em \lower 4pt \hbox {$\sim$}}}
\def\ltapprox{\hbox{\raise 2pt \hbox {$<$} \kern-1.1em \lower 5pt \hbox 
{$\approx$}}}
\def\gtsim{\hbox{\raise 2pt \hbox {$>$} \kern-1.1em \lower 4pt \hbox {$\sim$}}}
\def\gtapprox{\hbox{\raise 2pt \hbox {$>$} \kern-1.1em \lower 5pt \hbox 
{$\approx$}}}

\def\arcsec{$^{\prime\prime}$}
\def\arcmin{$^{\prime}$}
\def\degrees{$^{\circ}$}

\begin{document}
   \title{A2255: the First Detection of Filamentary Polarized Emission in a Radio Halo}


   \author{F. Govoni\inst{1,2}
          \and
          M. Murgia\inst{1,3}
          \and
          L. Feretti\inst{1}
         \and
          G. Giovannini\inst{1,2}
          \and
          D. Dallacasa\inst{1,2}
         \and
          G. B. Taylor\inst{4,5}
          }

   \offprints{F. Govoni, \email{fgovoni@ira.cnr.it}}

   \institute{Istituto di Radioastronomia -- CNR/INAF, 
              via Gobetti 101, I--40129 Bologna, Italy
           \and     
              Dipartimento di Astronomia, 
              Univ. Bologna, Via Ranzani 1, I--40127 Bologna, Italy
         \and
              INAF - Osservatorio Astronomico di Cagliari,
              Poggio dei Pini, Strada 54, I--09012 Capoterra (CA), Italy 
          \and
              National Radio Astronomy Observatory, Socorro, NM 87801, USA
          \and
              Kavli Institute of Particle Astrophysics and Cosmology,
              Menlo Park, CA 94025, USA
              }

   \date{Received ; accepted}

   \abstract{
A deep radio observation of the A2255 cluster of galaxies
has been carried out at 1.4 GHz with the
Very Large Array synthesis telescope.
Thanks to the excellent ($u$,$v$) coverage and sensitivity achieved by our 
observation, the low brightness diffuse extended sources 
in the cluster (radio halo and relic) have been imaged 
with unprecedented resolution and dynamic range. 
We find that the radio halo has filamentary structures 
that are strongly polarized.
The fractional linear polarization reaches levels of $\simeq$ 20$-$40\%
and the magnetic fields appear ordered on scales of $\sim$400 kpc.
This is the first successful attempt 
to detect polarized emission from a radio halo and provides 
strong evidence that in 
this cluster
the magnetic field is ordered on large scales.

   \keywords{Galaxies: clusters :
general -- Galaxies: intergalactic medium -- Polarization -- Magnetic fields -- Radio continuum:
general --
               }
   }

   \maketitle
%

\section{Introduction}

\begin{figure*}
\centering
\includegraphics[width=10 cm]{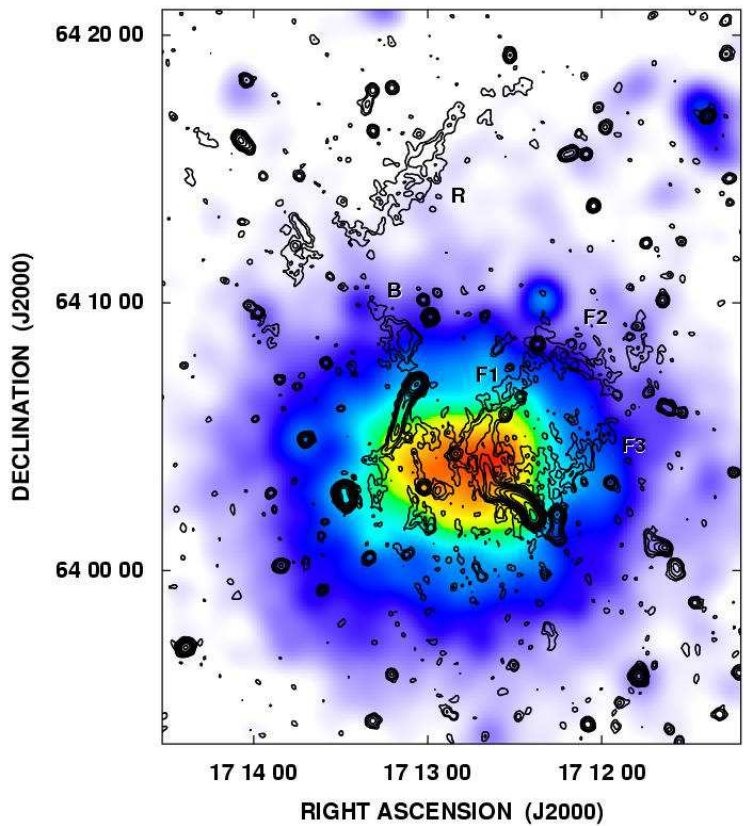}
\caption{Radio contours of the cluster A2255 
overlaid on the Rosat X-ray image (colors).
The radio image is at 1.4 GHz and has a FWHM of 
15\arcsec $\times$ 15\arcsec (uniform weighting). 
The sensitivity (1$\sigma$) is 16 $\mu$Jy/beam and the dynamic range 
is $\simeq$ 6300.
Contour levels are: 0.05 0.1 0.2 0.4 0.8 1.6 3.2 6.4 12.8 25.6 51.2 mJy/beam.
No primary beam correction has been applied to the image.
}
\label{fig1}
\end{figure*}

An ever increasing number of galaxy clusters exhibit
large-scale, diffuse, steep-spectrum
synchrotron sources associated with the intracluster medium.
These radio sources have been generally
classified as radio halos or relics depending on their
morphology and location.
Radio halos are located at the cluster center 
and are characterized by a
regular shape and extremely low surface brightness.
Relics are similar, but are found at the cluster periphery  
and in general have an elongated shape.
In some clusters, both a central halo and a peripheral 
relic are present.
While relics are usually strongly polarized,
no significant polarization has been detected so far in radio halos.
In the Coma cluster, the upper limit to the halo fractional polarization
is $\sim$ 10\% at 1400 MHz (Feretti \& Giovannini 1998).
Good upper limits ($<$ 5\%) have been placed for the powerful
radio halos of the galaxy clusters A2219, A2163 and $1E0657-57$ 
(Bacchi et al. 2003, Feretti et al. 2001, Liang et al. 2000). 

In addition to the analysis of the wide, diffuse synchrotron sources,
cluster magnetic fields can be constrained
through the detection 
of non-thermal emission of inverse
Compton origin in the hard X-ray wavelengths or
by studying Faraday rotation measure (RM) of polarized radio galaxies.
It is known that these different 
methods of analysis give somewhat discrepant results for the magnetic field
strength 
(see e.g. reviews by Carilli \& Taylor 2002, 
Govoni \& Feretti 2004, and references therein).
The knowledge of the magnetic field structure may be the key issue 
to understand the origin of this discrepancy.
Recently, En{\ss}lin \& Vogt (2003), Vogt \& En{\ss}lin (2003), 
Murgia et al. (2004), showed that the RM 
of radio galaxies can be used to infer 
not only the cluster magnetic field strength, 
but also the power spectrum of the cluster magnetic 
field fluctuations.
Murgia et al. (2004) derived the magnetic field power spectrum 
of a sample of clusters for which good RM data of cluster 
radio galaxies were
available. They found that A2255 appears as one of the clusters with a 
very steep intracluster magnetic field power spectra.
Moreover they pointed out that
morphology and polarization information of radio halos may
provide important constraints on the 
power spectrum of the magnetic field fluctuations 
on large scales.
In particular, their simulations showed that if the intracluster magnetic 
field fluctuates up to scales of some hundred kpc,
then steep magnetic field power spectra 
may give rise to detectable polarized filaments.

A2255 is a nearby (z=0.0806, Struble \& Rood 1999), 
rich cluster which shows signs of 
undergoing a merger event.
It is characterized by the presence of
a diffuse radio halo source at the cluster center,
a relic source at the cluster periphery, and several 
embedded head-tail radio galaxies (Jaffe \& Rudnick 1979,
Harris et al. 1980, Burns et al. 1995, Feretti et al. 1997).

We observed A2255 with the purposes
of detecting polarized emission
from the radio halo and obtaining information on the degree of
ordering of the cluster magnetic field.
In this letter we report the results of this high resolution and 
sensitivity observation.

Throughout this paper we assume a $\Lambda$CDM cosmology with
$H_0$ = 71 km s$^{-1}$Mpc$^{-1}$,
$\Omega_m$ = 0.3, and $\Omega_{\Lambda}$ = 0.7.
At the distance of A2255, 1\arcsec~ corresponds to 1.5 kpc. 

\section{Radio Data}
\begin{figure*}
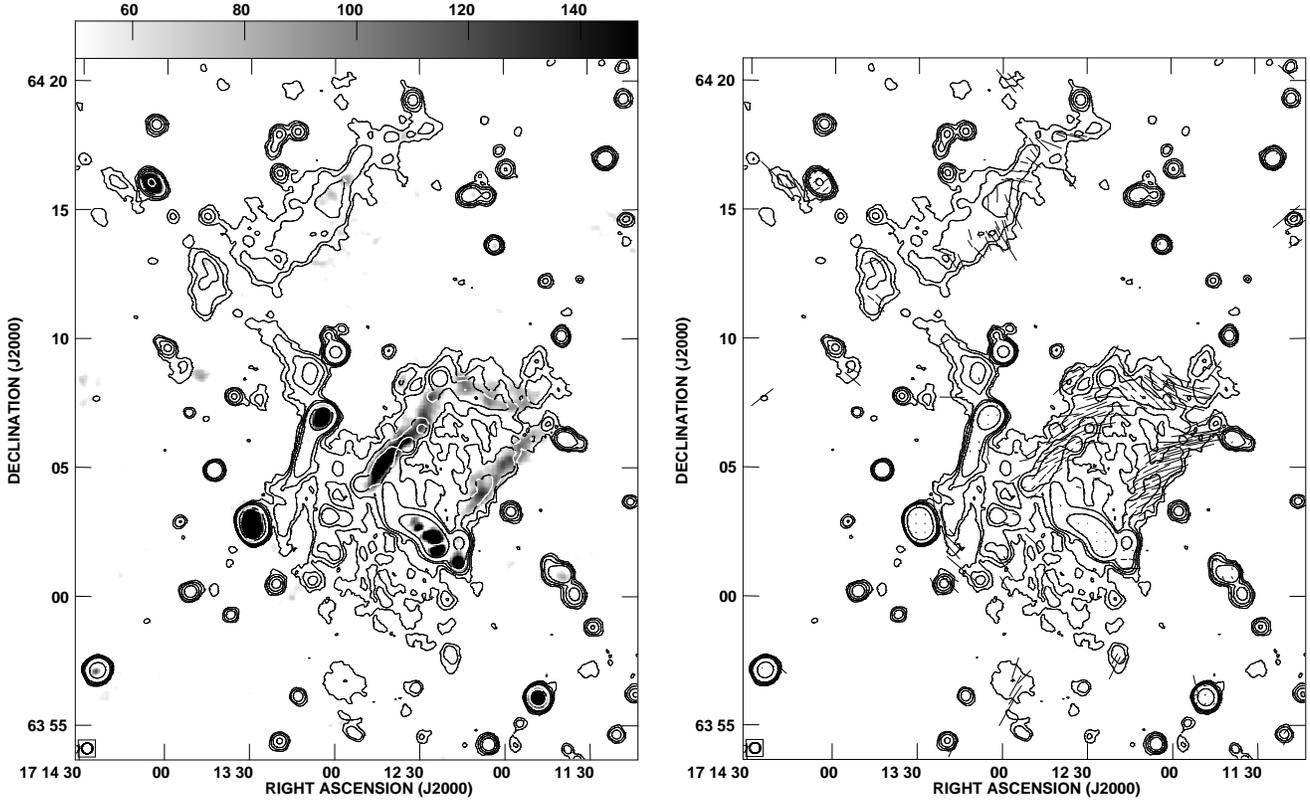

\centering
\includegraphics[width= 8.8 cm]{fig2a.ps}
\includegraphics[width= 8.8 cm]{fig2b.ps}
\caption{Total intensity radio contours of A2255 at 1.4 GHz with a
FWHM of 25\arcsec$\times$25\arcsec (natural weighting). 
The sensitivity (1$\sigma$) is $\simeq$ 24 $\mu$Jy/beam and the dynamic 
range is $\simeq$ 6000.
Contour levels are: 0.07 0.14 0.28 0.56 4.48 mJy/beam.
No primary beam correction has been applied.
Left: the contours of the total intensity 
are overlaid on the linearly 
polarized intensity (grey-scale). The sensitivity (1$\sigma$)
of the U and Q images is $\simeq$ 11 $\mu$Jy/beam.
The grey-scale shows the polarized flux from 50-150 $\mu$Jy/beam.
Right: the contours of the total intensity 
are overlaid on the polarization vectors.
The vector orientation represents the projected E-field
(not corrected for the contribution of the galactic rotation)
while their length is proportional to the fractional polarization
(1\arcmin=50\%).
All pixels with a fractional polarization less than 2$\sigma$, 
or with an error in the 
polarization angle greater than 10\degrees, have been blanked.
}
\label{fig2}
\end{figure*}

We observed A2255   
with the Very Large Array 
(VLA) in C configuration on 2004 April 19,
for a total integration time of $\simeq$11 hours on source.
This full-synthesis produced excellent ($u$,$v$) coverage. 
A bandwidth of 25 MHz was used for each of the two IF channels centered at  
1465 MHz and 1415 MHz.
The observing frequencies were selected in order to 
avoid interferences. 
Faraday rotation effects 
between the two channels should be small for RMs $<$100 rad m$^{-2}$.
The choice of the bandwidth was selected to reduce bandwidth smearing
effects.
Calibration and imaging were performed with the 
NRAO\footnote{The National Radio Astronomy Observatory (NRAO) 
is a facility of the National Science Foundation, operated under 
cooperative agreement by Associated Universities, Inc.}  
Astronomical Image Processing System
(AIPS), following the standard procedure:
Fourier-Transform, Clean and
Restore. Self-calibration was applied to remove residual 
phase variations.
Images in the Stokes parameters I, Q and U were made, 
using uniform and natural weighting.
The images of the polarized intensity (corrected for the
positive bias), the
fractional polarization and the position angle of polarization
were derived from the I, Q and U images.
The final images, shown here, were convolved  with a 
circular Gaussian with a FWHM of 15\arcsec~ and 25\arcsec~ respectively.    

\section{Results}
A full resolution (FWHM=15\arcsec) image of A2255, 
including several radio galaxies, the central halo
and the peripheral relic, is shown in Fig. 1 (contours).
The radio emission is overlaid on the Rosat X-ray image (Feretti et al. 1997)
to show the gas distribution of the cluster.
The diffuse extended radio sources are resolved and
at this high resolution only their brighter regions are visible.
The relic (label R) 
appears as a long, straight filament of about 12\arcmin~
in length located 10\arcmin~ to the North-East
from the cluster center and elongated in the 
South-East North-West direction. 
The halo shows a complex morphology.
The most prominent features detected at this resolution
are straight filaments (labels F1, F2, F3),
each about 6\arcmin~ in length and 2\arcmin~ in width,
nearly perpendicular to each other.
A bridge of low brightness emission (label B),
seems to connect the halo and the relic on the northeast side.
The filaments F1 and F3 are almost parallel to the relic,
while the bridge B is nearly parallel to the filament F2. 

In Fig. 2 (contours) a 
lower resolution (FWHM=25\arcsec)
image of the cluster radio emission is shown.
Owing to the higher signal to noise ratio,
the low brightness regions of the
diffuse sources are easily visible.
On the left, the contours of the total intensity 
are overlaid on the linear 
polarized intensity (grey-scale). 
On the right, the contours of the total intensity
are overlaid on the polarization vectors.
The vector orientation shows the projected E-field and their 
length is proportional to the fractional polarization
(1\arcmin=50\%).
In the figure, all pixels
with a fractional polarization less than 2$\sigma$  
or with an error in the polarization angle greater than 10\degrees~
were blanked.
The polarized emission at 15\arcsec~ resolution (not shown here) 
displays similar results. 
The most important result is that the bright filaments of the
halo appear strongly polarized at levels of $\simeq$ 20$-$40\%
($\simeq$4$\sigma$$-$8$\sigma$ detections). 
Regions of ordered magnetic field of $\sim$400 kpc in size
can be observed.
In the rest of the cluster we don't detect significant 
polarized emission except in the brighter regions of the relic
where the fractional polarization is in the 
range $\simeq$15$-$30\% ($\simeq$3$\sigma$$-$7$\sigma$ detections). 
The upper limit (2$\sigma$) to the fractional polarization in the 
fainter regions of the halo (i.e. where the average total intensity emission 
is about 0.15 mJy/beam), is $\simeq$ 15\%. 
The galactic RM in the direction of A2255 
is expected to be about $-6$ rad m$^{-2}$, based on the average
of the RM galactic contribution published by
Simard-Normandin et al. (1981) for sources near the cluster.
 Therefore, even if no Faraday rotation occurs within the cluster, 
the position angle of the E-field observed at 1.4\,GHz 
is rotated by $\sim$15\degrees~ counter-clockwise
with respect to the intrinsic (at $\lambda$=0) orientation.
The electric polarization vectors of the relic tend to 
be roughly perpendicular to the relic elongation
indicating aligned magnetic field structures within it,
while the electric polarization vectors of the halo seems roughly 
parallel to the filaments.

Fig. 3 shows the total intensity 
image at 25\arcsec~ resolution, 
with the discrete sources
subtracted.
The discrete sources were identified by making an image
using long spacings, then their components were subtracted 
directly in the {\it uv}-plane (AIPS task UVSUB).
To estimate the flux density of the cluster diffuse emission
the primary beam correction was applied to the image in Fig. 3
(AIPS task PBCOR).
The halo has a total flux density of $\simeq$ 56$\pm$3 mJy, 
the relic  $\simeq$ 23$\pm$1 mJy,
and their connecting bridge  $\simeq$ 6$\pm$0.5 mJy.
The three filaments F1, F2, F3 have flux densities of
9$\pm$0.5, 3$\pm$0.5, and 5$\pm$0.5mJy respectively, indicating
a total flux $\simeq$30\% of the flux of the entire halo.

\section{Discussion}

The absence of a significant polarization in halos
has been interpreted as the result of two concurrent effects:
internal Faraday rotation and beam depolarization.
The thermal intracluster gas is mixed with the relativistic
plasma, thus due to internal Faraday rotation,
significant depolarization may occur within the radio halos.
Moreover as a consequence of their 
extremely low surface brightness, 
radio halos have been studied so 
far at low spatial resolution. This could result
in a significant decrease
of the observed fractional polarization, if the cluster
magnetic field is tangled on scales smaller than the beam.

Murgia et al. (2004) showed that if the outer
scale of the magnetic field fluctuations extends up to
some hundred kpc, and if the power spectrum
\footnote{$\rm{|B_k|^2\propto k^{-n}}$ where n 
is the index of the power spectrum of the magnetic field fluctuations. 
The power spectrum 
is expressed as a vectorial form in $k$-space.} 
of the cluster magnetic field is relatively steep (n$\geq$3) 
there could be a chance
of detecting filamentary polarized emission in the halo.  
The deep and high resolution radio observation 
of A2255 presented here confirms their prediction.
The radio halo of A2255 shows, for the first time,
filaments of strong polarized emission.
Moreover, the distribution of the polarization angles
indicates that the magnetic field of this cluster
fluctuates up to scales of about 400 kpc in size.

The detection of polarized emission in a synchrotron halo in A2255 opens up 
new questions regarding its nature, origin, and connection 
with the history of merging.
The halo filaments could result from a
compression wave, which enhances and aligns disordered
magnetic fields. 
Most turbulence theories involve the processes by which the energy
is injected into a medium at large spatial scales and than converted into 
motions at smaller and smaller spatial scales until reaching 
scales at which it is dissipated.
What we have detected in A2255 may be 
the injection in the intracluster medium of energy on large scales, 
produced for example by a shock during a cluster merger.
However both the radio morphology of the filaments nearly perpendicular 
to each other, and the electric polarization vectors running
roughly parallel to the elongation of the radio halo filaments 
(indicating magnetic field structures perpendicular within them) 
are quite  unusual and difficult to explain in this framework.
Spatial spectral index information, in conjunction with high resolution
cluster X-ray and temperature images, will test
electron re-acceleration models (e.g. turbulence, shock)
responsible for the halo filaments formation and could help to determine whether
these structures are produced by shock waves resulting from 
a cluster merger.

Another important issue is to understand why the halo filaments are so 
strongly polarized.
They could be cluster foreground structures in which
the Faraday rotation is negligible.
RM information, obtained at well separated frequencies,
will be of great importance to evaluate this possibility.
In this case they should have a low rotation measure.

Finally it is not clear whether these filamentary polarized structures,
are typical features of clusters or if 
A2255 is a peculiar case.
One can devise a scenario in which all clusters have
magnetic fields fluctuating both on
small (visible thorough RM data) and large scales
(visible thorough radio halos). But only for those 
clusters for which the power spectrum of the magnetic field 
fluctuations is steep enough will these polarized 
filamentary structures be detectable.
Future observations on other clusters containing radio halos, 
and selected on the basis of their magnetic field power spectrum,
are necessary to test these ideas.

\begin{figure}
\centering
\includegraphics[width= 8.5 cm]{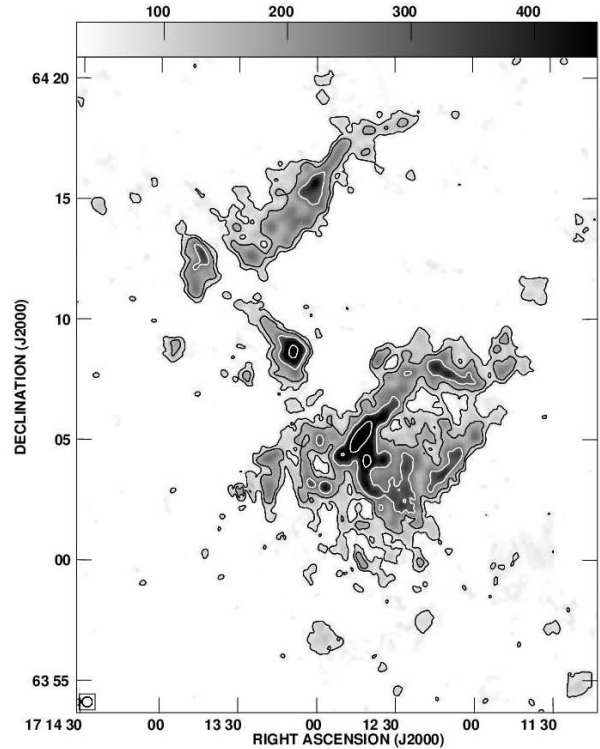}
\caption{
The 1.4 GHz image of the cluster radio emission
after subtraction of the discrete sources.
The FWHM is 25\arcsec $\times$ 25\arcsec (natural weighting). 
Contour levels are: 0.07 0.14 0.28 0.56 1.12 2.24 4.48 8.96 19.92 35.84 
71.68 mJy/beam.
The grey scale flux is $30-450$ $\mu$Jy/beam.
No primary beam correction has been applied to this image.
}
\label{fig3}
\end{figure}



\end{document}